\documentclass{article}

\input{tcilatex}

\begin{document}

{\footnotesize {Mathematical Problems of Computer Science {\small 25, 2006,
53--56.}}}

\bigskip 

\bigskip

\bigskip 

\begin{center}
{\Large \textbf{Interval Colourings of Some Regular Graphs}}

{\normalsize Rafael R. Kamalian and Petros A. Petrosyan}

{\small Institute for Informatics and Automation Problems of NAS of RA}

{\small e-mails rrkamalian@yahoo.com, pet\_petros@yahoo.com}

\bigskip

\textbf{Abstract}
\end{center}

A lower bound is obtained for the greatest possible number of colors in an
interval colourings of some regular graphs.

\bigskip

Let $G=(V,E)$ be an undirected graph without loops and multiple edges [1], $%
V(G)$ and $E(G)$ be the sets of vertices and edges of $G$, respectively. The
degree of a vertex $x\in V(G)$ is denoted by $d_{G}(x)$, the maximum degree
of a vertex of $G$-by $\Delta (G)$, and the chromatic index [2] of $G$-by $%
\chi ^{\prime }(G)$. A graph is regular, if all its vertices have the same
degree. If $\alpha $ is a proper edge colouring of the graph $G$ [3], then
the color of an edge $e\in E(G)$ in the colouring $\alpha $ is denoted by $%
\alpha (e,G)$, and by $\alpha (e)$ if from the context it is clear to which
graph it refers. For a proper edge colouring $\alpha $, the set of colors of
the edges that are incident to a vertex $x\in V(G)$, is denoted by $%
S(x,\alpha )$.

A proper colouring $\alpha $ of edges of $G$ with colors $1,2,\ldots ,t$ is
interval [4], if for each color $i,1\leq i\leq t,$ there exists at least one
edge $e_{i}\in E(G)$ with $\alpha (e_{i})=i$ and the edges incident with
each vertex $x\in V(G)$ are colored by $d_{G}(x)$ consecutive colors.

For $t\geq 1$ let $\mathcal{N}_{t}$ denote the set of graphs which have an
interval $t$-colouring, and assume: $\mathcal{N}\equiv
\dbigcup\limits_{t\geq 1}$ $\mathcal{N}_{t}$. For $G\in \mathcal{N}$ the
least and the greatest values of$\ t$, for which $G\in \mathcal{N}_{t}$, is
denoted by $w(G)$ and $W(G)$, respectively.

In [5] it is proved:

\textbf{Theorem 1}. Let $G$ be a regular graph.

\QTP{Body Math}
1) $G\in \mathcal{N}$ iff $\chi ^{\prime }(G)=\Delta (G)$.

\QTP{Body Math}
2) If $G\in \mathcal{N}$ and $\Delta (G)\leq t\leq W(G)$, then $G\in 
\mathcal{N}_{t}$.

\QTP{Body Math}
[6] and theorem 1 imply that for regular graphs the problem of deciding
whether $G\in \mathcal{N}$ or $G\notin \mathcal{N}$, is $NP$-complete [7,8].

\QTP{Body Math}
In this paper we will consider regular graphs $G=(V,E)$, where

\begin{center}
$V(G)=\left\{ x_{j}^{(i)}|\text{ }1\leq i\leq k,1\leq j\leq n\right\} $,

$E(G)=\left\{ \left( x_{p}^{(i)},x_{q}^{(i+1)}\right) |\text{ }1\leq i\leq
k-1,1\leq p\leq n,1\leq q\leq n\right\} \bigcup $

$\bigcup \left\{ \left( x_{p}^{(k)},x_{q}^{(1)}\right) |\text{ }1\leq p\leq
n,1\leq q\leq n\right\} $, $k\geq 3$.
\end{center}

It is not hard to see that $\Delta (G)=2n$. Let $\ \mathcal{R}\left(
n,k\right) $ be the set of all those graphs.

In [9] it is shown that if $G\in \mathcal{R}\left( n,k\right) $ then

\begin{center}
$\bigskip \chi ^{\prime }(G)=\left\{ 
\begin{array}{lll}
2n, & \text{if} & n\cdot k\text{ is even,} \\ 
2n+1, & \text{if} & n\cdot k\text{ is odd.}%
\end{array}%
\right. $
\end{center}

Theorem 1\textbf{\ }implies:

\QTP{Body Math}
\textbf{Corollary 1}. Let $G\in \mathcal{R}\left( n,k\right) $ . Then:

\QTP{Body Math}
1) $G\in \mathcal{N}$, if $n\cdot k$-is even;

\QTP{Body Math}
2) $G\notin \mathcal{N}$, if $n\cdot k$-is odd.

\QTP{Body Math}
\textbf{Corollary 2}. If $G\in \mathcal{R}\left( n,k\right) $ and $n\cdot k$%
-is even, then $w(G)=2n$.

\QTP{Body Math}
\textbf{Theorem 2}. If $G\in \mathcal{R}\left( n,k\right) $ and $k$-is even,
then $W(G)\geq 2n+\frac{n\cdot k}{2}-1$.

\textbf{Proof}. Let

\begin{center}
$V(G)=\left\{ x_{j}^{(i)}|1\leq i\leq k,1\leq j\leq n\right\} $,

$E(G)=\left\{ \left( x_{p}^{(i)},x_{q}^{(i+1)}\right) |\text{ }1\leq i\leq
k-1,1\leq p\leq n,1\leq q\leq n\right\} \bigcup $

$\bigcup \left\{ \left( x_{p}^{(k)},x_{q}^{(1)}\right) |\text{ }1\leq p\leq
n,1\leq q\leq n\right\} $.
\end{center}

Let $G_{1}$ be the subgraph of the graph $G$ induced by the vertices $%
x_{1}^{(k)},x_{2}^{(k)},...,x_{n}^{(k)}$, $%
x_{1}^{(1)},x_{2}^{(1)},...,x_{n}^{(1)}$. It is clear that $G_{1}$ is a
regular complete bipartite graph of the degree $n$. Therefore $\chi ^{\prime
}(G_{1})=\Delta (G_{1})=n$, and due to theorem 1, $G_{1}\in \mathcal{N}$.

Consider a proper edge colouring $\alpha $ of the graph $G_{1}$ defined as
follows:

\begin{center}
$\alpha \left( \left( x_{p}^{(k)},x_{q}^{(1)}\right) \right) =p+q-1$ for $%
p=1,2,\ldots ,n$ and $q=1,2,\ldots ,n$.
\end{center}

It is not hard to check that $\alpha $ is an interval $\left( 2n-1\right) $%
-colouring of the graph $G_{1}$.

Define an edge colouring $\beta $ of the graph $G$ in the following way:

1) $\beta \left( \left( x_{p}^{(k)},x_{q}^{(1)}\right) ,G\right) =\alpha
\left( \left( x_{p}^{(k)},x_{q}^{(1)}\right) ,G_{1}\right) $\bigskip

for $p=1,2,\ldots ,n$ and $q=1,2,\ldots ,n$;

\bigskip 2) for $i=1,2,\ldots ,\frac{k}{2}-1$ and $p=1,2,\ldots ,n$, $%
q=1,2,\ldots ,n$

\begin{center}
$\beta \left( \left( x_{p}^{(i)},x_{q}^{(i+1)}\right) ,G\right) =\beta
\left( \left( x_{p}^{(k-i)},x_{q}^{(k-i+1)}\right) ,G\right) =\alpha \left(
\left( x_{p}^{(k)},x_{q}^{(1)}\right) ,G_{1}\right) +i\cdot n$;
\end{center}

\bigskip 3) $\ \beta \left( \left( x_{p}^{(\frac{k}{2})},x_{q}^{(\frac{k}{2}%
+1)}\right) ,G\right) =\alpha \left( \left( x_{p}^{(k)},x_{q}^{(1)}\right)
,G_{1}\right) +\frac{n\cdot k}{2}$

for$\ \ p=1,2,\ldots ,n$ and $q=1,2,\ldots ,n$.

Let us show that $\beta $ is an interval $\left( 2n+\frac{n\cdot k}{2}%
-1\right) $-colouring of the graph $G$.

First of all note that for $i,$ $1\leq i\leq 2n-1$ there is an edge $%
e_{i}\in E(G)$ such that $\beta (e_{i})=i$.

Now let us show that for $j,$ $2n\leq j\leq 2n+\frac{n\cdot k}{2}-1$ there
is an edge $e_{j}\in E(G)$ with $\beta (e_{j})=j$.

Consider the vertices $x_{n}^{(2)},x_{n}^{(3)},...,x_{n}^{(\frac{k}{2})}$.
The definition of $\beta $ implies that

\begin{center}
$\dbigcup\limits_{i=2}^{\frac{k}{2}}S\left( x_{n}^{(i)},\beta \right)
=\left\{ 2n,2n+1,...,2n+\frac{n\cdot k}{2}-1\right\} $.
\end{center}

This proves that for $j,$ $2n\leq j\leq 2n+\frac{n\cdot k}{2}-1$ there is an
edge $e_{j}\in E(G)$ with $\beta (e_{j})=j$.

Let us show that the edges that are incident to a vertex $\ v\in V(G)$ are
colored by $2n$ consecutive colors.

Let $x_{j}^{(i)}\in V(G)$, $1\leq i\leq k,1\leq j\leq n$.

Case 1: $i=1,1\leq j\leq n$.

The definition of $\beta $ implies that:

\begin{center}
$S\left( x_{j}^{(1)},\beta \right) =\left( \dbigcup\limits_{l=1}^{n}\beta
\left( \left( x_{l}^{(k)},x_{j}^{(1)}\right) \right) \right) \bigcup \left(
\dbigcup\limits_{m=1}^{n}\beta \left( \left( x_{j}^{(1)},x_{m}^{(2)}\right)
\right) \right) =$

$=\left( \dbigcup\limits_{l=1}^{n}\alpha \left( \left(
x_{l}^{(k)},x_{j}^{(1)}\right) ,G_{1}\right) \right) \bigcup \left(
\dbigcup\limits_{m=1}^{n}\left( \alpha \left( \left(
x_{j}^{(k)},x_{m}^{(1)}\right) ,G_{1}\right) +n\right) \right) =$

$=\left\{ j,j+1,...,j+n-1\right\} \cup \left\{ j+n,j+n+1,...,j+2n-1\right\}
=\left\{ j,j+1,...,j+2n-1\right\} $.
\end{center}

Case 2: $2\leq i\leq k-1,1\leq j\leq n$.

The definition of $\beta $ implies that:

\begin{center}
$S\left( x_{j}^{(i)},\beta \right) =\left( \dbigcup\limits_{l=1}^{n}\beta
\left( \left( x_{l}^{(i-1)},x_{j}^{(i)}\right) \right) \right) \bigcup
\left( \dbigcup\limits_{m=1}^{n}\beta \left( \left(
x_{j}^{(i)},x_{m}^{(i+1)}\right) \right) \right) =$

$=\left( \dbigcup\limits_{l=1}^{n}\left( \alpha \left( \left(
x_{l}^{(k)},x_{j}^{(1)}\right) ,G_{1}\right) +\left( i-1\right) \cdot
n\right) \right) \bigcup \left( \dbigcup\limits_{m=1}^{n}\left( \alpha
\left( \left( x_{j}^{(k)},x_{m}^{(1)}\right) ,G_{1}\right) +i\cdot n\right)
\right) =$

$=\left\{ j+(i-1)\cdot n,...,j+i\cdot n-1\right\} \cup \left\{ j+i\cdot
n,...,j+(i+1)\cdot n-1\right\} =$

$=\left\{ j+(i-1)\cdot n,...,j+(i+1)\cdot n-1\right\} $.
\end{center}

Case 3: $i=k,1\leq j\leq n$.

The definition of $\beta $ implies that:

\begin{center}
$S\left( x_{j}^{(k)},\beta \right) =\left( \dbigcup\limits_{l=1}^{n}\beta
\left( \left( x_{l}^{(k-1)},x_{j}^{(k)}\right) \right) \right) \bigcup
\left( \dbigcup\limits_{m=1}^{n}\beta \left( \left(
x_{j}^{(k)},x_{m}^{(1)}\right) \right) \right) =$

$=\left( \dbigcup\limits_{l=1}^{n}\left( \alpha \left( \left(
x_{l}^{(k)},x_{j}^{(1)}\right) ,G_{1}\right) +n\right) \right) \bigcup
\left( \dbigcup\limits_{m=1}^{n}\alpha \left( \left(
x_{j}^{(k)},x_{m}^{(1)}\right) ,G_{1}\right) \right) =$

$=\left\{ j+n,j+n+1,...,j+2n-1\right\} \cup \left\{ j,j+1,...,j+n-1\right\}
=\left\{ j,j+1,...,j+2n-1\right\} $.
\end{center}

\textbf{Theorem 2\ }is proved.

\textbf{Corollary\ 3.} If $G\in \mathcal{R}(n,k)$, $k$-is even and $2n\leq
t\leq 2n+\frac{n\cdot k}{2}-1$, then $G\in \mathcal{N}_{t}$.

Let us note that if $G\in \mathcal{R}(n,4)$, then the lower bound of the
proved theorem is the exact value of $W(G)$, that is $W(G)=4n-1$.

$\bigskip $

\begin{center}
\bigskip
\end{center}

\end{document}